# Structural Polymorphism of the Cytoskeleton:

# A Model of Linker-Assisted Filament Aggregation

Itamar Borukhov[1,2,*], Robijn F. Bruinsma[2],
William M. Gelbart[1] and Andrea J. Liu[1,3]

[1] Department of Chemistry and Biochemistry
University of California, Los Angeles, CA 90095, USA

[2] Department of Physics and Astronomy
University of California, Los Angeles, CA 90095, USA

[3] Department of Physics and Astronomy
University of Pennsylvania, Philadelphia, PA 19104, USA

Corresponding Author:  Andrea J. Liu, Department of Physics & Astronomy, University of Pennsylvania, 209 S. 33rd St., Philadelphia, PA 19104, USA; 215-573-7374 (office), 215-898-2010 (fax), ajliu@physics.upenn.edu.

Classification:  Physical Sciences, Physics

---

[*] Current address: Compugen Ltd., 72 Pinchas Rosen  St., Tel Aviv 69512 Israel, email: itamar01@gmail.com

## Abstract

The phase behavior of charged rods in the presence of inter-rod linkers is studied theoretically as a model for the equilibrium behavior underlying the organization of actin filaments by linker proteins in the cytoskeleton. The presence of linkers in the solution modifies the effective inter-rod interaction and can lead to inter-filament attraction. Depending on the system's composition and physical properties such as linker binding energies, filaments will either orient perpendicular or parallel to each other, leading to network-like or bundled structures. We show that such a system can have one of three generic phase diagrams, one dominated by bundles, another by networks, and the third containing both bundle and network-like phases. The first two diagrams can be found over a wide range of interaction energies, while the third occurs only for a narrow range. These results provide theoretical understanding of the classification of linker proteins as bundling proteins or crosslinking proteins. In addition, they suggest possible mechanisms by which the cell may control cytoskeletal morphology.

## I. Introduction

Filamentous actin (F-actin) is a highly-charged, stiff biopolymer that is abundant in the cell and a key component of the cellular cytoskeleton. A cell controls the assembly of actin filaments into structures ranging from dilute networks, where filaments cross at large angles, to dense bundles, where filaments are closely-packed and nearly parallel to one another [1]. This *structural polymorphism* of actin filaments is crucial to cell function because different structures play different roles. Actin bundles play a key role during cell locomotion and cell adhesion [1]. On the other hand, actin networks near the periphery of an animal cell form the cell cortex, which controls the mechanical properties of the cell surface. To exploit the functional differences between bundles and networks, cells switch between formation of these two structures during cell crawling and cell division [2,3].

Regulation of actin architecture requires control of both the kinetics of assembly and disassembly and of the morphology of actin aggregates. The assembly and disassembly of actin structures are carefully regulated by a number of specific actin-binding proteins such as branching, capping and severing proteins [1]. On the other hand, the morphology, structure and stability of actin aggregates are controlled by *linker proteins*, architectural proteins that can bind two filaments together [4,5]. These are typically classified into *bundling* proteins, primarily found in bundles, and *crosslinking* proteins, primarily found in networks. In addition, multivalent cationic species such as $Ca^{2+}$, $Mg^{2+}$ and spermine can control the morphology of actin aggregation and should therefore also be regarded as linkers.

In this paper, we focus on the equilibrium phase behavior of actin/linker systems, with a view towards understanding structural polymorphism in the cytoskeleton. It is generally believed that the phase diagram of actin/linkers is of the generic sol/gel type [2,6,7] described by percolation models [8], with bundle formation as a separate phenomenon occurring at high linker concentration. In contrast, we propose that the morphology of actin/linker aggregates is inherently polymorphic due to the existence of several different thermodynamic phases, with bundle formation an integral part of the phase diagram. The novel ingredient of the actin/linker system, as compared to conventional cross-linked polymers, is the importance of *competing interactions*, which are well-known to produce exceptionally rich phase diagrams [9,10,11,12]. In actin/linker systems, electrostatic repulsion between filaments and possible linker preference favor large-angle crossings between filaments. On the other hand, steric repulsions are minimized and linker binding is maximized when filaments are parallel; these effects favor parallel orientation. We find four competing phases, shown in Fig. 1. The first one is the *isotropic* (I) phase, where filaments are randomly oriented; this maximizes orientational entropy. The second one is the *nematic* (N) phase, where filaments preferentially orient in one direction; this increases translational entropy at the expense of orientational entropy. The remaining two phases are produced by the linker-filament interaction. The first of these is the *cubatic* (C) phase [13,14], which has long-ranged orientational order in three mutually-perpendicular directions, but no translational order. Thus, the cubatic phase is a network phase with cubic *orientational* order but the rods or linker junctions are not organized into a periodic lattice. It appears when two filaments prefer to orient perpendicularly, either because a specific linker prefers perpendicular alignment, or because of electrostatic repulsion between the filaments. The resulting 90° junctions organize into a structure with cubic orientational symmetry, which maximizes the number of right-angle crossings[*]. Finally, the *bundle* (B) phase is crystalline with rods parallel to each other and organized in a hexagonal array, with linkers binding them together along their lengths. The nematic and bundle phases look similar but in the nematic phase the inter-filament interaction remains repulsive despite the presence of linkers, while in the bundle phase it is attractive.

A cubatic phase, with true long-ranged orientational order, is highly unlikely to be observed in a real system due to kinetic constraints associated with the difficulty for linker-junctions to reorganize once they are formed. However, if the cubatic were the equilibrium phase, we would expect to observe disordered networks of single filaments crossing each other at large angles, with linkers at the junctions. Similarly, one would not expect to find a single macroscopic bundle, but rather isolated bundles or networks of bundles, depending on concentration. Our key claim is that the equilibrium transition between the cubatic and bundle phases underlies structural switching between networks and bundles in the actin/linker system.

---

[*] We do not consider the possibility of a translationally-ordered cubic phase because the spacing between filaments in the cubatic phase is large compared with the interaction range. Thus, translational order would cost entropy with no gain in energy and should therefore be unfavorable with respect to the cubatic. We note, however, that the cubatic is merely the simplest representative of a larger class of tetratic phases [14] where the symmetry between the three principal directions is broken. An example for such a phase is the squaratic phase, a special case of the biaxial nematic where the two orientational axes are perpendicular. Close to the isotropic/cubatic transition, the cubatic phase is favorable relative to the squaratic because of the higher number and orientational entropy of $90^0$ junctions in the cubatic phase. We note, however, that at higher rod concentrations the squaratic phase or even a cubic phase might be stable. Our aim here is to treat the cubatic phase as a typical representative of the network-like tetratic phases that are rich in $90^o$ junctions. Since the free energies in the different tetratic phases can be quite close to each other a detailed calculation is needed to distinguish between them. Including such details in the theory should not change our qualitative conclusions and will only obscure the overall picture.

## II. Model and Method

In our system, actin filaments are replaced by perfectly rigid, charged rods [*][15] of length $L$ and diameter $D \ll L$. The complexity of all the different linker proteins is reduced to a few physical properties - their size and their binding energies. Because linkers prefer to bind to two rods and therefore tend to bring rods together, they induce an effective attraction between rods. Instead of addressing the two-component rod-linker mixture directly, we eliminate the linkers as an explicit species by allowing the rods to interact via effective potentials that depend on linker concentration. We employ the virial approximation, a variant of the Onsager theory of the isotropic/nematic transition of uncharged rigid rods without linkers [16], to study the statistical mechanics of a solution of rods with these exotic potentials. Because the rods are long, transitions between phases occur at low concentrations [16]; this justifies truncation of the virial approximation at the second term. As a result, we need never consider linker-mediated interactions of more than two rods at a time. This approach allows us to calculate the phase diagram.

### A. Effective rod-rod interaction

The first step in the analysis is to determine the effective rod-rod interaction free energy in the presence of the linkers. The interaction free energy $u(r,\gamma)$ depends on the rod-rod separation, $r$, and the angle between the two rods, $\gamma$. The full potential $u(r,\gamma)$ could be obtained from detailed molecular dynamics simulations for a given linker protein. Such simulations have been carried out for model rods for the case where the linker is a polyvalent counterion [17,18]. Here we choose an alternate approach to gain insight into trends with binding energy, etc. We construct a simple model potential which captures the main features of any realistic potential (see Fig. 2):

---

[*] For filament lengths exceeding the persistence length (of order $10\mu m$ for F-actin), one must replace $L$ with the persistence length [15].

$$u(r,\gamma) = \begin{cases} \infty & r \leq D \\ u(\gamma) + \Gamma(\gamma) & D < r \leq D + \delta \\ \Gamma(\gamma)e^{-\kappa(r-D-\delta)} & D + \delta < r \end{cases} \quad (1)$$

Three different interactions contribute to the inter-rod potential: (i) hard-core repulsions dominate at short distances $(r \leq D)$. (ii) Linker-mediated attractions are present at intermediate ranges $(D < r \leq D + \delta)$ and can be written as

$$u(\gamma) = \begin{cases} \dfrac{L}{D}u_{\parallel} & \gamma \leq \dfrac{D}{L} \\ \dfrac{u_{\perp}}{\sin\gamma} & \gamma > \dfrac{D}{L} \end{cases} \quad (2)$$

where $u_{\parallel}, u_{\perp}$ characterize the longer-ranged repulsion at small and large angles, respectively and $\delta$ is of the order of the linker size. Finally, (iii) screened electrostatic repulsions compete with the linker-mediated attractions at intermediate distances and dominate the long-range behavior $(r > D + \delta)$. Here,

$$\Gamma(\gamma) = \begin{cases} \dfrac{L}{D}\Gamma_{\parallel} & \gamma \leq \dfrac{D}{L} \\ \dfrac{\Gamma_{\perp}}{\sin\gamma} & \gamma > \dfrac{D}{L} \end{cases} \quad (3)$$

where $\Gamma_{\parallel}$ and $\Gamma_{\perp}$ characterize the longer-ranged repulsion at small and large angles, respectively [19]. Recent simulations for the case where the linkers are multivalent counterions [18] show that this form (Fig. 2) does indeed provide a reasonable description of the effective potential.

The model potential $u(r,\gamma)$ is characterized by four energies (see Eqs. (1)-(3)). These energies govern the strengths of the linker-mediated attractions, $u_{\parallel}$ and $u_{\perp}$, and of the screened electrostatic repulsions, $\Gamma_{\parallel}$ and $\Gamma_{\perp}$. Because both the attraction per linker and the repulsion per linker depend on the angle $\gamma$ between the rods, we allow for separate quantities, distinguished by the subscripts ∥ and ⊥ to characterize the interactions at small and large angles, respectively. There is also an additional factor of $L/D$ that multiplies $u_{\parallel}$ and $\Gamma_{\parallel}$, but not $u_{\perp}$ and $\Gamma_{\perp}$, in the interaction potential in Fig. 2. This reflects the property that the interaction between two parallel rods is proportional to their length, $L$, while the interaction between two rods crossing at large angles is independent of $L$ provided the interaction range is smaller than $L$.

The repulsion energies $\Gamma_{\parallel}$ and $\Gamma_{\perp}$ depend on the charge on the two rods and the screening length. Within Debye-Hückel theory,

$$\Gamma_{\parallel} = 2D(\lambda/e)^2 \ell_B K_0 \left[\kappa(D+\delta)\right] / \left[(\kappa D/2)K_1(\kappa D/2)\right]^2$$
$$\Gamma_{\perp} = 2\pi\kappa^{-1}(\lambda/e)^2 \ell_B e^{-\kappa(D+\delta)} / \left[(\kappa D/2)K_1(\kappa D/2)\right] \quad (4)$$

where $\lambda$ is the linear charge density along the rods, $\kappa^{-1}$ is the screening length, and $K_0$ and $K_1$ are the zero and first order Bessel functions, respectively. All energies are expressed in units of the thermal energy $k_B T$. The Bjerrum length, defined as $\ell_B = e^2 / \varepsilon k_B T$, is about 7Å for an aqueous solution at room temperature; here, $e$ is the unit charge and $\varepsilon$ is the dielectric constant of the solution. Under physiological conditions, $\kappa^{-1}$ is about 1nm so the entire interaction depicted in Fig. 2 is quite short-ranged.

The free energy scales characterizing the linker-mediated attractions, $u_\parallel$ and $u_\perp$, depend on the linker concentration (or alternatively, the linker chemical potential $\mu_l$) and binding energies. Linkers can be in one of three states: (i) free in solution, (ii) adsorbed to a single filament with adsorption energy $\varepsilon_1 < 0$, or (iii) bound to two filaments with a binding energy $2\varepsilon_\parallel < 0$ when the rods are parallel to each other or $2\varepsilon_\perp < 0$ when the rods cross at large angles. The resulting linker-induced attractions can be estimated using a simple Langmuir adsorption model:

$$
\begin{aligned}
(L/D)u_\parallel &= -n_\parallel \ln\left[1 + e^{\mu_l - 2\varepsilon_\parallel}\right] + 2n_\parallel \ln\left[1 + e^{\mu_l - \varepsilon_1}\right] \\
u_\perp &= -n_\perp \ln\left[1 + e^{\mu_l - 2\varepsilon_\perp}\right] + 2n_\perp \ln\left[1 + e^{\mu_l - \varepsilon_1}\right]
\end{aligned}
\tag{5}
$$

The first term represents the contribution of linkers bound to the two filaments. Here $n_\parallel = L/\delta$ and $n_\perp = 1$ are the number of binding sites per filament when the rods are parallel and perpendicular, respectively; $\delta$ being the linker diameter. The second term represents the reference free energy when the rods are far apart; it is double the free energy of linkers adsorbed to a single filament. The contribution of free linkers cancels out but will be included later on in the free energy.

From the two-rod interaction potential shown in Fig. 2, one can already see the competing interactions that determine whether the cubatic phase or the bundle phase is stable. Both $u_\parallel$ and $u_\perp$ depend on the linker concentration through the linker chemical potential $\mu_l$ (Eq. (5)), and can be either positive or negative. As a result, the interaction $u(\gamma)$ can *change sign* as a function of linker chemical potential. At $\gamma = 0°$, $u(\gamma) = L/D(\Gamma_\parallel - |u_\parallel|)$ (denoted as I in Fig. 2), and at $\gamma = 90°$, $u(\gamma) = \Gamma_\perp - |u_\perp|$ (denoted as II in Fig. 2). If, as the linker chemical potential increases, $L/D(\Gamma_\parallel - |u_\parallel|)$ becomes negative before $\Gamma_\perp - |u_\perp|$, then the system will tend to bundle. If $\Gamma_\perp - |u_\perp|$ becomes negative first, then the system will tend to form a cubatic phase (network).

## B. The generalized Onsager theory

To calculate the equilibrium phase diagram, allowing for the isotropic (I), nematic (N), cubatic (C) and bundle (B) phases, we adopt the approach of Onsager [16]. We write the free energy per unit volume of a solution of rods in terms of the rod orientational distribution function, $f(\Omega)$, up through quadratic order in the rod concentration:

$$
F = c_r(\ln c_r - 1) + c_r \sigma(\{f\}) + \frac{1}{2}c_r^2 w(\{f\}) + f_{\text{linkers}}
\tag{6}
$$

Here the dimensionless rod concentration is defined as $c_r = (\pi/4)L^2 D_{\text{eff}}\rho_r$ where $\rho_r$ is the number of rods per unit volume, $D_{\text{eff}} = D + \kappa^{-1}(c_E + \ln\Gamma_\perp)$ is the effective rod diameter and $c_E$ is Euler's constant. The effective diameter $D_{eff}$ can be interpreted as the distance at which the interaction between two screened charged rods is comparable to the thermal energy. The first three terms are those included by Onsager. The first term is the contribution from the translational entropy of the rods, the second term is from the orientational entropy, with $\sigma(\{f\}) = \int d\Omega f(\Omega)\ln[4\pi f(\Omega)]$, and the third term is the contribution from the two-body interaction, with $w(\{f\}) = \int d\Omega_1 d\Omega_2 f(\Omega_1)f(\Omega_2)B(\gamma_{12})$ and $B(\gamma_{12}) = \int dr_{12}[1 - \exp(-u(r_{12}, \gamma_{12}))]$. The effect of linkers enters in two ways, through the effective potential $u$ in the second virial coefficient, and through the final term in Eq. (6):

$$f_{\text{linkers}} = -\frac{L^2 D_{\text{eff}}}{v_l}\exp(\mu_l) - n_{\parallel}c_r \log[1 + \exp(\mu_l - \varepsilon_1)] \tag{7}$$

The first term in Eq. (7) is the contribution of free linkers in the solution and is due to their translational (ideal gas) entropy. The second term is the contribution of linkers adsorbed to isolated rods that comprises the reference terms that were subtracted from Eq. (5).

Given the inter-rod potential embodied in Fig. 2, we can calculate the second virial coefficient $B(\gamma)$ that determines the functional $w(\{f\})$ in Eq. (6). We find

$$B(\gamma) = \begin{cases} 8\dfrac{D^2 + \kappa^{-2}\ln^2(\Gamma_\parallel L/D)}{D_{\text{eff}}L} + 16\dfrac{D\delta}{D_{\text{eff}}L}\left[1 + \dfrac{\exp[-(|u_\parallel| - \Gamma_\parallel)L/D] - 1}{(|u_\parallel| - \Gamma_\parallel)L/D}\right] & \gamma \leq D/L \\ \dfrac{8}{\pi}\sin\gamma\left[1 - \dfrac{\kappa^{-1}}{D_{\text{eff}}}\ln\sin\gamma\right] + \dfrac{8\delta}{\pi D_{\text{eff}}}\sin\gamma\left[1 - \exp\left(-\dfrac{\Gamma_\perp}{\sin\gamma} + |u_\perp|\right)\right] & \gamma > D/L \end{cases} \tag{8}$$

In both cases, the first term arises from the hard-core and screened Coulomb repulsions, while the second term arises from the linker-mediated attraction. Our theory is only valid when $B(\gamma)$ is positive; for negative $B(\gamma)$, the free energy is unstable and higher-order virial terms must be included. It is clear from Eq. (8) that the second term changes sign at small angles when $L/D(\Gamma_\parallel - |u_\parallel|)$ changes sign (see I in Fig. 2), and changes sign at large angles when $\Gamma_\perp - |u_\perp|$ does (see II in Fig. 2). As a result, $B(0°)$ becomes negative when $L/D(\Gamma_\parallel - |u_\parallel|)$ is negative and of order unity, while $B(90°)$ changes sign when $\Gamma_\perp - |u_\perp|$ is negative and of order unity.

The structure of the phase diagram is determined by whether $B(0°)$ or $B(90°)$ first becomes negative as the linker chemical potential increases. When $B(0°)$ becomes negative before $B(90°)$, the phase diagram is dominated by bundles; in the opposite case, it is dominated by the cubatic phase (networks). If $B(90°)$ and $B(0°)$ change sign at approximately the same linker chemical potential, then networks can coexist with bundles.

In principle, one could construct the phase diagram by minimizing the free energy in Eq. (6) with respect to the orientational distribution function, $f(\Omega)$. However, it suffices to postulate a particular form for $f(\Omega)$ in each phase with a single variational parameter that characterizes its width, and then to compare the resulting free energies for each phase. In the isotropic phase, $f(\Omega) = 1/4\pi$ since all orientations are equally likely. In a nematic phase oriented in the $z$-direction, $f(\Omega)$ is peaked at $\theta = 0°$ and $\theta = 180°$, and one can characterize the distribution by the width of the peaks. We have adopted a *cone approximation*, which assumes a uniform distribution of rod orientations within a narrow cone of width $\Delta\theta \ll 1$, with no possibility of orientations outside the cone. The nematic phase is characterized by a pair of cones, one at $\theta = 0°$ and the other at $\theta = 180°$ (see Fig. 1), while the cubatic phase is characterized by three pairs of mutually perpendicular cones (see Fig. 1). Given Eq. (8) for the second virial coefficient, $B(\gamma)$, we can use the cone approximation to obtain analytical expressions for the orientational entropy $\sigma$ and the integrated two-body term $w$ in Eq. (6) for the isotropic, nematic and cubatic phases. For a fixed linker chemical potential, coexistence between any pair of phases is then determined by equating the rod chemical potentials and osmotic pressures. The linker concentration in each phase is then calculated from the free energy and linker chemical potential, yielding coexistence curves in the $c_r$, $c_l$ plane, as shown in Fig. 3.

The filament length $L$ determines the rod concentrations at which the different phase transitions occur. The concentrations characterizing the isotropic/nematic transition [16] and isotropic/cubatic transition both scale as $1/L^2 D_{\mathrm{eff}}$, where $D_{\mathrm{eff}}$ is the effective rod diameter. This corresponds to low volume fractions of order $D_{\mathrm{eff}}/L$ and therefore justifies the use of the second virial approximation and explains why our choice of $c_r = (\pi/4)L^2 D_{\mathrm{eff}}\rho_r$ as the dimensionless rod concentration. Note that we adopt the same scaling for the linker concentration: $c_l = (\pi/4)L^2 D_{\mathrm{eff}}\rho_l$.

We treat the bundle phase by assuming a dense hexagonal packing of rods and highly restricted orientational distribution function with $\Delta\theta = D/L$ (note that in contrast to the nematic phase, $\Delta\theta$ vanishes for bundles in the limit of infinite rod length). In the calculations discussed below, we use the following lengths, which were chosen to represent a solution of actin filaments with 100mM monovalent salt. The rod diameter is $D = 8$ nm, the rod length is $L = 1000$ nm, the size of the binding protein (or equivalently, range of the linker-mediated attraction) is $\delta = 1$ nm, and the screening length is $\kappa^{-1} = 1$ nm. The energies that characterize the electrostatic repulsion are calculated from Eq. (4) to be $\Gamma_\parallel = \Gamma_\perp = 1.26$ and the binding energy of a linker to a single filament is set equal to $\varepsilon_l = -7$ (recall that energies are measured in units of the thermal energy $k_B T$, so $\varepsilon_l = -7$ corresponds to a binding energy of –4 kcal/mol).

# III. Results

Depending on the linker binding energies, $\varepsilon_\parallel$ and $\varepsilon_\perp$, we find that a solution of charged rods with linkers must have one of only three qualitatively-different phase diagrams. These three phase diagrams are depicted in Fig. 3. Nearly all of the binding energy parameter space leads to one of the two phase diagrams depicted in Fig. 3(a) and (c). For $\varepsilon_\perp$ comparable to or less negative than $\varepsilon_\parallel$, we find the "bundle-dominated" phase diagram in Fig. 3(a), where there are bundles but no cubatic phase (no networks). For $\varepsilon_\perp$ somewhat more negative than $\varepsilon_\parallel$, on the other hand, we find the "network-dominated" phase diagram shown in Fig. 3(c), where there are networks (the cubatic) but no bundles.

In Fig. 3(a) the two binding energies are the same: $\varepsilon_\parallel = \varepsilon_\perp = -7$. When the concentration of linkers is low, the standard first order isotropic/nematic phase transition is recovered. The coexistence tie lines (not shown) are tilted upward on the right, reflecting a higher concentration of linkers in the nematic phase than in the isotropic phase. The phase boundaries appear vertical because linkers do not play a significant role in the isotropic/nematic phase transition, which is driven by repulsive rod-rod interactions. As the concentration of linkers increases, the collective adsorption of many linkers between pairs of rods becomes strong enough to change the sign of the rod-rod interaction. We then reach three different coexistence regions. At low rod concentrations, a linker-poor isotropic phase coexists with a linker-rich bundle phase while at high rod concentrations, a linker-poor nematic phase coexists with a linker-rich bundle phase. At intermediate rod concentrations, we encounter a three-phase region, where isotropic, nematic and bundle phases coexist[*]. Note that the linker concentration at the onset of bundling increases approximately linearly with rod concentration; this reflects the property that as more rods are added, correspondingly more linkers are needed to induce bundling [20,21]. The slope of this line increases with linker binding energy.

If $\varepsilon_\perp \ll \varepsilon_\parallel < 0$, then the phase diagram is dramatically reorganized as shown in Fig. 3(c), where $\varepsilon_\perp = -12$ (-7 kcal/mol) while $\varepsilon_\parallel = -7$ is unchanged. As before, at low linker concentrations there is an isotropic phase and a nematic phase separated by a coexistence region. Note that the vertical scales are different so this region is barely visible in Fig. 3(c). In this case, the coexistence tie lines are tilted upwards on the left so that the linker concentration is higher in the isotropic phase, because there are more large-angle crossings in the isotropic phase than in the nematic phase. At higher linker concentrations, we find a cubatic phase that can coexist with the isotropic phase, the nematic phase, or both. The isotropic phase is bounded by an instability line where the second virial coefficient $B(90°)$ defined in Eq. (8) becomes negative (shaded region). Our theory breaks down in the shaded region.

---

[*] The bundle phase itself is not shown in Fig. 3(a) because the linker and rod concentrations in the bundle phase are much higher than those in the coexisting isotropic and nematic phases. This is because in the isotropic, nematic and cubatic phases the linker/rod ratio is typically of order unity while in the bundle phase it is of order $L/\delta$.

For intermediate values of $\varepsilon_\perp$, we find the third phase diagram, shown in Fig. 3(b), where $\varepsilon_\perp = -10.5$. Unlike the other two, this diagram contains *both* the cubic phase and the bundle phase, with the former at low linker concentration and the latter at higher linker concentrations. For binding energies large compared to $k_BT$, this intermediate phase diagram is remarkably difficult to find; it exists only over a very narrow range of values of $\varepsilon_\perp$. By increasing $\varepsilon_\perp$ (or decreasing $\varepsilon_\parallel$) by a fraction of $k_BT$, one can jump straight from the bundle-dominated phase diagram (Fig. 3(a)) to the network-dominated one (Fig. 3(c)). The range encompassed by this intermediate phase diagram is small because many linkers can bind if the filaments are parallel while only a few can bind if they are perpendicular. This magnification of the binding energy when filaments are parallel enables a slight change of binding energy to tip the system from networks to bundles.

In the case where the binding energies are of order $k_BT$, the range encompassed by the intermediate phase diagram, a fraction of $k_BT$, is not much smaller than the binding energies themselves. Thus, weakly-binding linkers such as multivalent cations, with binding energies of the same order as the thermal energy, are more likely to be described by the intermediate phase diagram.

# IV. Discussion

## A. *Comparison with* in vitro *experiments*

Phase diagrams have been measured for only a limited number of biopolymer/linker systems. The first of these is a solution of actin and polyethylene glycol (PEG) under high salt conditions, where an attractive interaction arises because of PEG-mediated depletion interactions while the electrostatic repulsion is small because of screening. For this case, $|\varepsilon_\parallel| >> |\varepsilon_\perp|$ [22] and the phase diagram shown in Fig. 3(a) should be realized. This is indeed observed by Suzuki, Yamazaki and Ito [23], who report a phase diagram that is qualitatively identical. In fact, even in the absence of PEG a linker-free actin solution should show this behavior at high salt, because van der Waals attractions also prefer parallel alignment [24]. Another system for which one would expect $|\varepsilon_\parallel| >> |\varepsilon_\perp|$ would be solutions of DNA or F-actin with multivalent species such as spermine or spermidine as linkers, since these elongated molecules are expected to have a preference for the parallel orientation. This is consistent with experimental observations by a number of groups. In particular, the linker concentration at the onset of bundling has been shown to increase linearly with rod concentration in DNA condensation experiments [20] and actin bundling [25].

For multivalent cations such as $Ca^{2+}$, $\varepsilon_\perp \approx \varepsilon_\parallel \approx -1$ should be satisfied. For such linkers, numerical simulations show that Minimum II at 90° in Fig. 2 goes negative before Minimum I at 0° with increasing linker chemical potential. In this case, the system might be described by Fig. 3(b), or by Fig. 3(a) with a metastable cubic phase. Wong, et al.

[26] have measured phase diagrams for actin/ $Ca^{2+}$ and actin/ $Mg^{2+}$ solutions. At high linker concentrations , they do observe bundles, consistent with Fig. 3(b). At low linker concentrations, they report two-dimensional network phases or "rafts," rather than the expected cubatic structure. Theoretical calculations [27] suggest that two-dimensional network structures may be kinetically-limited versions of the cubatic phase. However, the competition between rafts and the cubatic phase might also depend on the exact locations of linker binding sites on the filaments as well as on properties of the linkers, such as their shape and flexibility.

A complementary calculation by Zilman and Safran [7] allows for a distinction between junction-rich and junction-poor isotropic phases but does not include angle-dependent interactions. This approach suggests that for strong linkers, the entropy of inter-rod junctions can be a significant driving force for phase separation, favoring the bundle phase at the expense of an isotropic network.

To our knowledge, there are no detailed experimental measurements of phase diagrams corresponding to Fig. 3(c). Small ions with high positive oxidation states might yield such phase diagrams under low salt conditions.

## B. Bundling vs. crosslinking proteins

The scale of protein-protein interactions, and therefore of the binding energies between linker proteins and actin, is typically 10 $k_B T$ or more under physiological conditions. Therefore, linker proteins should yield phase diagrams Fig. 3(a) or Fig. 3(c), but only rarely Fig. 3(b). Indeed, linker proteins are usually classified as either "bundling" proteins or "crosslinking" proteins. Examples of the first group include fascin, scruin, villin and fimbrin, while examples of the second include filamin, spectrin and dystrophin. Our prediction is that bundling proteins should generically be described by the phase diagram in Fig. 3(a), while crosslinking proteins should be described by Fig. 3(c). Thus, our results provide a rationalization of the classification of linker proteins and imply that the functions of these proteins can be understood in terms of physical properties such as $\varepsilon_\perp$ and $\varepsilon_\parallel$. This suggests that atomistic simulations can yield insight into linker protein function. One could simulate a protein interacting with short segments of two actin filaments that can be placed parallel or perpendicular to each other, thereby estimating $\varepsilon_\parallel$ and $\varepsilon_\perp$.

Interestingly, for the case of α-actinin, both bundle and network-like structures have been reported [28,29]. This would indicate that the binding energy of α-actinin to actin should be unusually low, so that α-actinin might be described by Fig. 3(b).

The known result [1] that long linkers tend to lead to networks while short linkers tend to give rise to bundles arises naturally from our theory. From our expression for $B(\gamma)$, Eq. (8), it can be seen that the relative weight of the short-range linker-mediated attraction is proportional to the linker size, $\delta$. Thus, larger linkers exhibit stronger attractive contributions than smaller ones and can form single-linker junctions more easily. On the other hand, small linkers have weaker attractive contributions and can only form multi-linker bundles. Finally, we note that the screening length (roughly 1nm) determines the weight of the repulsive electrostatic interactions in $B(\gamma)$, so the ratio of the screening length to the linker size provides insight into the relative weight of these two competing interactions.

## C. Mechanisms for cellular control of structural polymorphism

The fact that a small change of binding energy can cause the system to switch from a bundle-dominated phase diagram to a network-dominated one suggests several ways in which the cell might control structural polymorphism. First, linker binding energies could be regulated by chemical modification. For example, it has been shown that phosphorylation of the linker protein VASP can affect its binding energy to actin and its bundling activity [30]. Alternatively, calcium binding can affect the binding energy of linker proteins to actin [31]. By regulating binding energies, the system could jump between the phase diagrams in Fig. 3(a) and Fig. 3(c).

Second, the addition of a small amount of a different linker protein could have a similar effect. In that case, the system can not only jump from one phase diagram to the other, but might develop new phases such as networks of bundles.

Third, regulation of filament length could affect phase behavior. As the rod length increases, the onset of the cubatic phase shifts to lower rod concentrations and the phase-coexistence regions narrow. Since the onset concentration scales as $1/L^2$, it is very sensitive to small changes of filament length. For example, in Fig. 3(c) the system could shift itself from isolated filaments (isotropic phase) to networks (cubatic) by increasing filament length [*].

Fourth, recall that *weakly*-binding proteins or multivalent cations such as $Ca^{2+}$ and $Mg^{2+}$ are more likely to be described by the intermediate phase diagram shown in Fig. 3(b). This may offer an interesting advantage for structural control, since there is an isotropic/cubatic and a cubatic/bundle phase transition with increasing linker concentration. The contrasting phase diagrams of weakly-binding and strongly-binding linkers suggests that the cell might use a two-step process to control morphology; for example, it might first tune the *concentration* of weakly-binding linker proteins to initiate bundle or network formation, then introduce strongly-binding bundling or crosslinking proteins to reinforce these structures [32].

---

[*] It should be noted that in the cell the filament length distribution is not fixed and is controlled (among other things) by the presence of linker proteins (Biron, D., Moses, E., Borukhov, I. & Safran, S. A., unpublished work).

Note that it is not necessary for the system to be in the coexistence region of the cubatic and bundle phases in order to see both networks and bundles. For long rods, the kinetic barriers to changing structural morphology are very large compared to the thermal energy. Suppose the system is initially described by the network-dominated phase diagram, and jumps to conditions described by the bundle-dominated one. From then on, any actin polymerization will lead to bundles, but the networks will remain until they are dismantled by depolymerization.

We thank K.-C. Lee, J. Theriot, E. Sackmann, G. Sowa, C. Safinya, H. Strey, G. Wong and A. Zilman for useful discussions. Support from NSF grants CHE-0096492 (IB and AJL), PHY99-07949 (IB) and Israel-U.S. BSF grant 97-00205 (IB and WMG) is gratefully acknowledged.

**Figure Captions**

Figure 1. The different phases included in our calculation and their corresponding rod orientational distributions. In the isotropic (I) phase, rods can be oriented in any direction, in the nematic (N) phase, rods tend to align in a preferential direction. In the cubic (C) phase, rods tend to align in three mutually-perpendicular directions and in the bundle (B) phase, rods are aligned and hexagonally close-packed.

Figure 2. The effective rod-rod potential described by Eqs. (1)-(3). This potential is characterized by a short-ranged repulsion, an intermediate-ranged attraction and a longer-ranged repulsion. It is characterized by four energies; $u_{\parallel} < 0$ and $u_{\perp} < 0$ represent the linker-mediated attractions and $\Gamma_{\parallel} > 0$ and $\Gamma_{\perp} > 0$ are the electrostatic repulsions when the rods are parallel and perpendicular, respectively. The minima at 0° and 90° can rise or fall depending on these four energies.

Figure 3. The three possible phase diagrams for solutions of charged rods and linkers. In all three cases, we use $\varepsilon_{\perp} = \varepsilon_{\parallel} = -7$. (a) Bundle-dominated phase diagram. Here $\varepsilon_{\perp} = -7$. Note that this phase diagram does not contain the cubatic phase anywhere. (b) Intermediate phase diagram. Here $\varepsilon_{\perp} = -10.5$. This diagram contains the cubatic phase at low linker concentration and the bundle phase at higher linker concentrations. (c) Network-dominated phase diagram. Here $\varepsilon_{\perp} = -12$. Note that this diagram does not contain bundles anywhere.

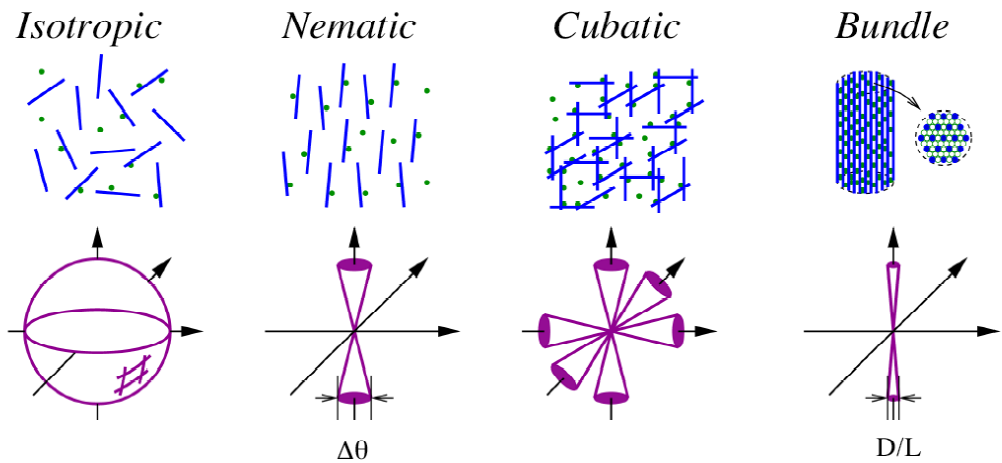

Figure 1, Borukhov et al.

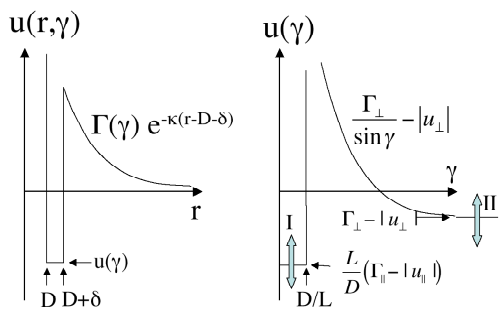

Figure 2, Borukhov et al.

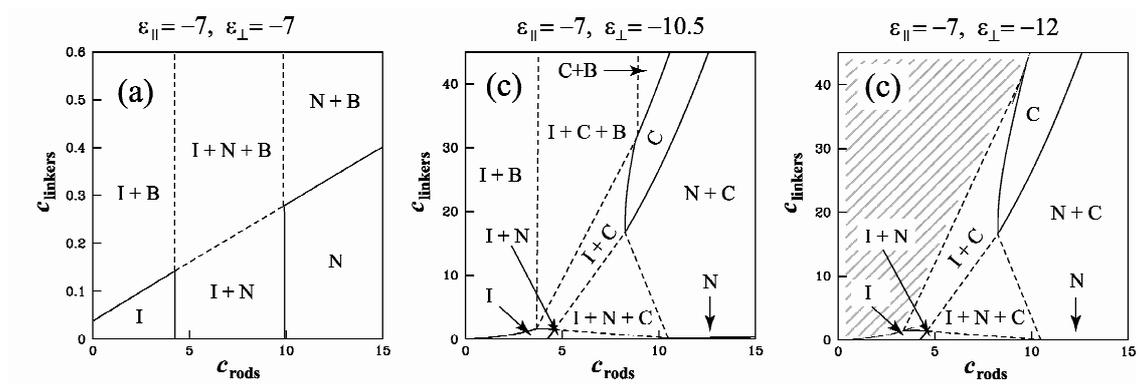

Figure 3, Borukhov et al.